\documentclass[11pt,a4paper,twocolumn]{article}

\usepackage[sort&compress,numbers]{natbib}

\usepackage{graphicx}
\usepackage{amsmath} 
\usepackage{amssymb} 

\usepackage{geometry} 
\usepackage[font={footnotesize}]{caption}
\usepackage{fancyhdr}

\usepackage{abstract}

\usepackage{titling}
\pretitle{\begin{flushleft}\LARGE\rmfamily\upshape\bfseries}
\posttitle{\par\end{flushleft}\vskip 0.5em}

\preauthor{\begin{flushleft}\large\rmfamily\scshape}
\postauthor{\par
        \vskip 0.5em        
        $^1$\textit{Department of Theoretical and Experimental Nuclear Physics, \\ Odessa  National Polytechnic University, Ukraine}

        $^2$\textit{Institute for Radiophysics and Electronics, NASU, Kharkov, Ukraine}

        $^3$\textit{Institute for Nuclear Research and Nuclear Energy, Sofia, Bulgaria}
\end{flushleft}}

\predate{\begin{flushleft}\large\scshape}
\postdate{\par\vskip -3em \end{flushleft}}

\usepackage[bookmarks=true,colorlinks=true,citecolor=blue,linkcolor=red,urlcolor=magenta]{hyperref}

\begin{document}


\title{Can Resonant Oscillations of the Earth Ionosphere Influence the Human Brain Biorhythm?}

\author{V.D.~Rusov$^1$\thanks{Cooresponding author e-mail: siiis@te.net.ua}, K.A.~Lukin$^2$, T.N.~Zelentsova$^1$, 
             E.P.~Linnik$^1$, M.E.~Beglaryan$^1$, V.P.~Smolyar$^1$, M.~Filippov$^1$ and B.~Vachev$^3$}


\fancyhead[LO]{V.D.~Rusov \textit{et al.} $\diamond$ "Pearls" and Human Brain Biorhythm}
\fancyhead[RO]{\thepage}
\fancyhead[RE]{V.D.~Rusov \textit{et al.}}
\fancyhead[LE]{\thepage}
\fancyfoot[C]{}

\date{}


\twocolumn[
  \maketitle             
  \thispagestyle{empty}
%
%
%
%
  \begin{onecolabstract} 
    Within the frames of Alfv\'{e}n sweep maser theory the description of morphological features of geomagnetic pulsations in the ionosphere with frequencies (0.1-10~Hz)  in the  vicinity of Schumann resonance (7.83~Hz) is obtained. It is shown that the related regular spectral shapes of geomagnetic pulsations in the ionosphere determined by "viscosity" and "elasticity" of magneto-plasma medium that control the nonlinear relaxation of energy and deviation of Alfv\'{e}n wave energy around its equilibrium value. Due to the fact that the frequency bands of Alfv\'{e}n maser resonant structures practically coincide with the frequency band \textit{delta}- and partially \textit{theta}-rhythms of human brain, the problem of degree of possible impact of electromagnetic "pearl" type resonant structures (0.1-5~Hz) onto the brain bio-rhythms stability is discussed.
    
    
  \end{onecolabstract}
   ]
\saythanks            

\section{Introduction}
\label{intro}

Lately Nobelist L. Montagnier's group has published three articles deeply challenging the standard views about genetic code and providing strong support for the notion of water memory \citep{ref1,ref2,ref3}. In a series of delicate experiments \citep{ref1,ref2} they demonstrated the possibility of the emission of low-frequency electromagnetic Extremely Low Frequency (ELF) waves from bacterial DNA sequences and the apparent ability of these waves to organize nucleotides (the "building" material of DNA) into new bacterial DNA by mediation of structures within water \citep{ref4}.

Without going into details of physical justification of quantum-field interpretation of these results\footnote{New Scientist reacted by sharp article "Scorn over claim of teleported DNA" \citep{ref5}.}, let us emphasize one significant experimental result of this group. This result is related to stable detection of ELF waves ($<$7~Hz) from bacterial DNA sequences. Obviously, if the result is reproduced in similar experiments of another research groups, the unique importance of this fact is difficult to overestimate in understanding of living matter essence.

First of all, it concerns not only research of drastically new fundamental properties of spatial-temporal structure of eukaryotic genome, but also refers to studying of equally important issues that are related to exogenous nature of $<$7~Hz ELF waves impact  directly onto a human brain and its biorhythms. 

It is well known that the brain neurons constitute different types of networks that interact by means of electrical signals. Neuron networks configurations comprise electrical circuits of oscillatory type. Electrical oscillations with different frequencies correspond to different states of brain. These oscillations could be detected by brain electroencephalogram.

Numerous investigations have shown that electrical oscillations of different frequencies dominate in a healthy human being brain at its different states \citep{Ptitsyna1998,Kholodov1982}. Transitions between brain activities happen not continuously, but only in discrete steps, from one level to another. The rest state corresponds to the steadiest \textit{alpha}-rhythm with frequencies laying within the frequency band from 8 to 13~Hz. \textit{beta}-rhythm with boundary frequencies 14-35~Hz corresponds to brain work. The slowest oscillations at frequencies 0.5-4~Hz are typical for \textit{delta}-rhythm which corresponds to deep sleep. At last, \textit{theta}-rhythm with frequencies from 4 to 7~Hz dominates in the brain if the state of nuisance or danger appears.

At the same time it is known \citep{Binhi2002a,Binhi2003,Adair2000,Nittby2008,Santini2009} that the weak magnetic fields impact on biological systems is a subject of the biophysics section called magnetobiology. It studies the biological reactions and mechanisms of the weak fields action. Magnetobiology is a part of a general fundamental problem of the biological efficiency of the weak and ultraweak physicochemical factors, which operate below the biological defense mechanisms threshold, and so may be accumulated on a subcelluar level.

It is necessary to note here that there is no acceptable physical understanding of the way the weak magnetic fields cause the living systems reaction \citep{Binhi2003} so far, although it has been experimentally found that such fields may change the biochemical reactions rate sharply in a resonance-like way \citep{Binhi2003,Liboff2009}. The physical nature of this phenomenon is still unclear, and it forms one of the most important, if not a general, problem of magnetobiology which includes the co-called "$kT$ problem".

The problem consists in the fact that the weak magnetic field energy (say, geomagnetic filed) of the same order as the $kT$ heat energy is distributed over the volume 12 orders of magnitude larger (which approximately corresponds to a cell size). In such form the problem of the biological impact of the low-frequency magnetic oscillations has two aspects\citep{Binhi2003}:
\begin{itemize}
\item what is a mechanism of the weak low-frequency magnetic signal transformation that causes the changes on the biochemical processes level of $kT$ order?
\item what is a mechanism of such stability, i.e. how do such small impacts not get lost on the  heat disturbances of $kT$ order background?
\end{itemize}

Not going into details of this complex and fundamental problem, the essence of which is expounded in review by \citep{Binhi2003}, let us note that in spite of the stated magnetobiology difficulties, there are serious reasons to believe that the main features of the magnetobiological effect are reliably established in numerous experiments and tests and are reproducible on different experimental models and under different magnetic conditions. On the other hand, the answers to the above-mentioned questions "lie" in the nonequilibrium thermodynamics field. "It is generally known that metabolism in living systems is a combination of primary non-equilibrium processes. The origin and breakdown of biophysical structures at time smaller than the time of thermalization of all degrees of freedom in these structures provide a good example of systems that are far from equilibrium where even weak field quanta can be manifested in system's breakdown parameters. In other words, if the life (thermalization) time of certain degrees of freedom interacting with field quanta is larger than the system's characteristic time of life, then such degrees of freedom exist in the absence of temperature proper. Therefore, a comparison of their energy changes due to field quanta absorption with $kT$ has no sense" \citep{Binhi1997}. The candidates for the solution of this problem today are the mechanism of the molecule quantum states interference for the idealized protein cavity \citep{Binhi1997,Binhi2003} and the mechanism of the molecular gyroscope interference \citep{Binhi2002b,Binhi2003}.

Turning back to experiment, let us examine the possible physical causes of the low-frequency geomagnetic fields generation and the consequences of their impact on the eucaryotic cells in short.

\begin{figure}[h]
\begin{center}
  \includegraphics[width=6cm]{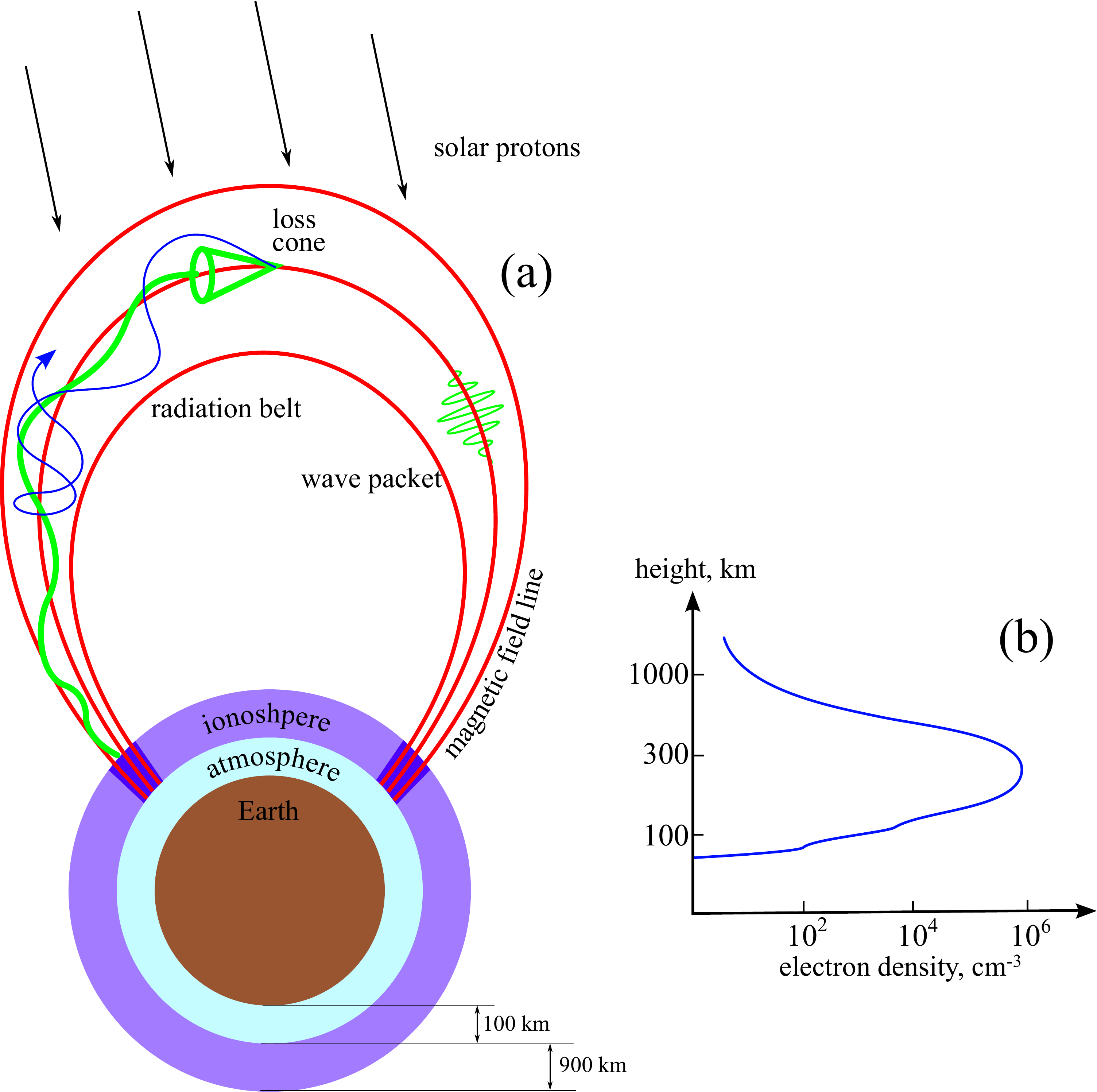}
\end{center}
\caption{Earth and surrounding electromagnetic resonant objects (adapted from \citep{ref6,ref7}). \textbf{a)} Air gap at the altitudes of 0-100~km is the global Schumann spherical resonator with 7.83~Hz resonant frequency; the altitude region of 100-1000~km is a dense ionized shell (ionosphere). Inside its mass ionosphere Alfv\'{e}n resonator with the first resonant frequency that varies in time within the limits of 0.5-3.0~Hz is located. Geomagnetic field lines lie above the ionosphere and are shown in red. High-energy protons cross this tubes. This Geomagnetic field line rests upon magneto-conjugated regions of the Earth's ionosphere which all together form the resonator of, so called, magnetosphere Alfv\'{e}n maser which generates "pearl" type electromagnetic signals. Traffic diagram for the particles in radiation belt is also depicted in the figure. Particles with velocities being inside the loss cone (green line), possess small transversal velocities and fall into dense layers of the atmosphere, whereas particles outside the loss cone (blue line with arrow) possess bigger transversal velocities and are captured by geomagnetic tube-trap due to their reflections from the magnetic mirrors of the ionosphere; \textbf{b)} Density of charged particles in plasma of Earth's ionosphere versus altitude.}
\label{fig1}
\end{figure}

It is well known  that Earth's atmosphere  between dense ionized shell called ionosphere (at an altitude of 100 km) and Earth's surface, possesses the electromagnetic resonant properties (Fig.~\ref{fig1}~\citep{ref6}). Hence, resonances of the spherical cavity "Earth's surface - ionosphere" manifest themselves in electromagnetic quasi-monochromatic signals that permanently present nearby Earth's surface and has certain impact onto the Earth's biosphere. Among resonances of this type in the frequency band between (0.1-10)~Hz the most known and studied is the so called Schumann resonance at the frequency of 7.83 Hz. This resonance is observed for electromagnetic waves with the wavelength exactly equal to the Earth's circle. Schumann resonance has drawn attention of physicians practically immediately after its discovery in connection with studying of impact of electromagnetic radiation onto the \textit{alpha}-rhythms of human brain which lie within the 8-13~Hz frequencies band.

\begin{figure}
\begin{center}
  \includegraphics[width=6cm]{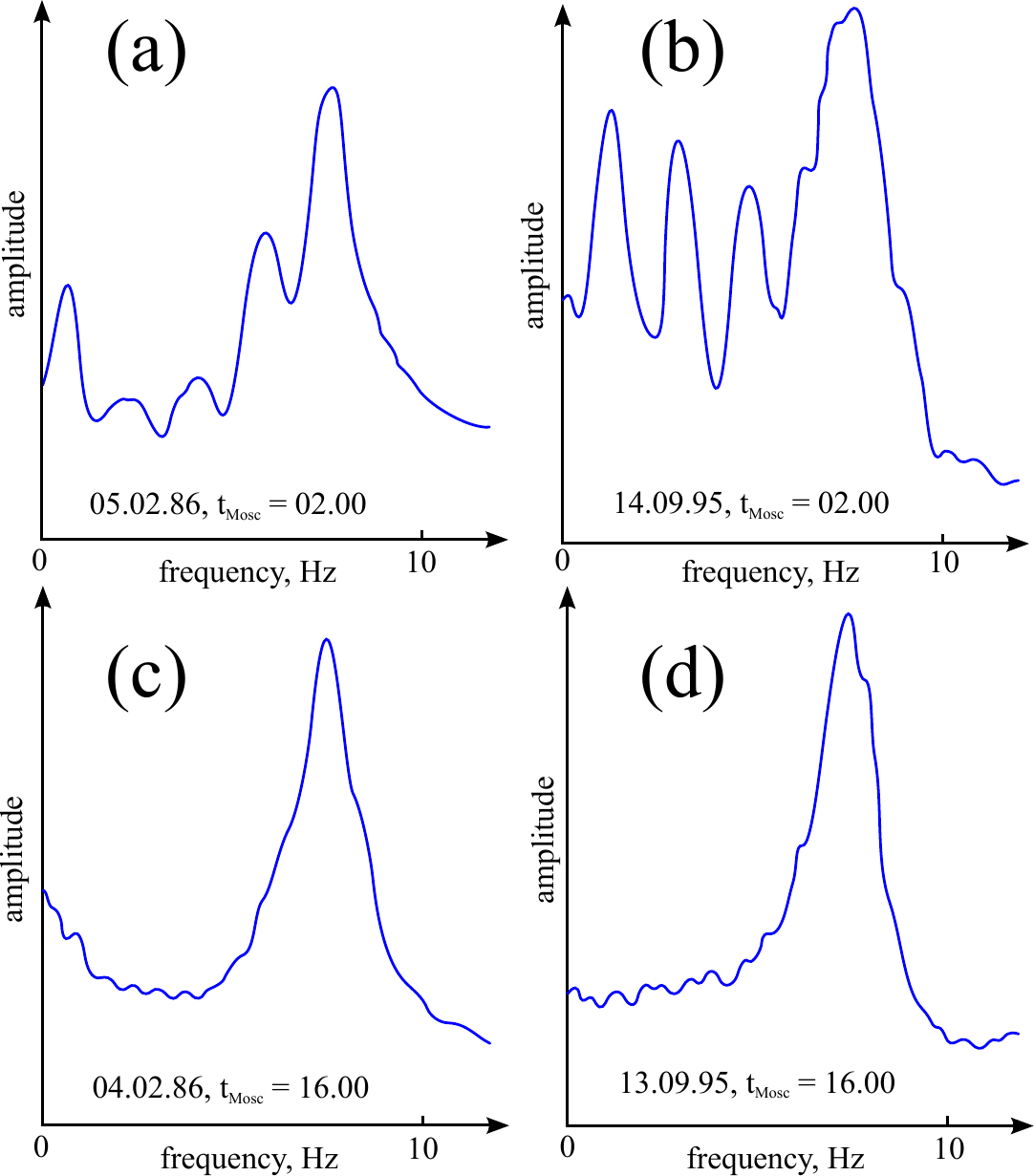}
\end{center}
\caption{Electromagnetic noise spectrum structure for middle latitudes has a pronounced resonant structure (adapted from~\citep{ref7}). In the daytime (\textbf{c)} and \textbf{d)}) the spectrum has a peak associated with Schuman resonance at 7.83~Hz, while at night time (\textbf{a)} and \textbf{b)}) electromagnetic noise produced by lightning's radiation at the frequencies below Schumann resonance is filtered out by Ionosphere Alfv\'{e}n Resonator (IAR). In the years of solar activity maximum the noise spectrum at night time is similar to that of daytime.}
\label{fig2}
\end{figure}

Thunderstorms feed the  Schumann resonator eternally. Initial frequency spectrum of electrical discharges (lightnings) during thunderstorms represents practically white noise. The resonant systems of near-Earth space filter out corresponding parts of the spectrum which is shown in Fig.~\ref{fig2}~\citep{ref7}.

Ionospheric Alfv\'{e}n Resonator (IAR) is being considered along with Schumann resonator  as a near-Earth resonant system, as well. In particular, with the help of IAR it was possible to explain new resonant radiation in the frequency band of 0.1-10~Hz~\citep{ref11}. This radiation was discovered in 1985 and characterized by quasi-periodic modulation (within frequency range of 0.5-3 Hz) of the oscillations. This modulation appears above the background noise electromagnetic spectrum of the atmosphere and has regular daily variation (Fig.~\ref{fig2}). It is not difficult to show \citep{ref11}, that within the frame of IAR model the resonant frequency $f_{res}$ of these oscillations is defined by ionosphere layer thickness $l$, Earth's magnetic field strength $H_{Earth}$, and concentration $n$ of particles with mass $M$
\begin{equation}
f_{res} = \frac{v_A}{2l}, ~~v_A = \frac{H_{Earth}}{\sqrt{4 \pi M n}},
\end{equation}

\noindent where $v_A$ is Alfv\'{e}n velocity. According to~\citep{ref11} the spectrum structure shown in Fig.~\ref{fig2} is defined by resonant frequency $f_{res}$ and its harmonics. For typical values of $H_{Earth} \sim 0.4~E$, $M \sim 1.5 \cdot 10^{-23}~g$, $n \sim 10^5 ~cm^{-3}$ and $l \sim 500~km$ this estimation gives $f_{res} \sim 2 ~Hz$. Apparently, this estimation is in a good agreement with experimental frequencies estimations of the detected radiations 0.5-3~Hz~\citep{ref11}.

Here it is interesting to mention another important role of IAR properties in affecting dynamics of larger scale resonator for Alfv\'{e}n waves: magnetospheric resonator for Alfv\'{e}n waves – Alfv\'{e}n Resonator (AR), formed by geomagnetic field line resting upon magneto-conjugated regions of Earth's surface. High-energy protons may cross geomagnetic field lines of the resonator and excite ultra-low frequency (ULF) electromagnetic oscillations practically in the same frequency band:  0.2-5 Hz , due to maser effect for the trapped protons and self-oscillatory mode of this resonator \citep{ref12}. This generator was called as magnetospheric Alfv\'{e}n maser \citep{ref6,ref7,ref8,ref9,ref10}, which is schematically shown in Fig.~\ref{fig1}. The signals generated via this mechanism are often referred to as "pearls". The spectral dynamical characteristics of the "pearls" and their temporal dependencies had been a mystery for researchers until recent times.  a remarkable fact had been discovered recently consisting in the strong negative correlation between intensity level of low-frequency resonant lines in the atmospheric noise radiation spectrum and solar activity~\citep{ref7}, and, correspondingly, between intensity  of resonant radiations of "pearl" type and solar activity  \citep{ref13,ref14,ref15}, that directly correlate with predictions for IAR models~\citep{ref7} and Alfv\'{e}n maser ones~\citep{ref10}.

Due to the fact that frequency bands of Alfv\'{e}n maser resonances practically coincide with the frequency band of \textit{delta}- and partially \textit{theta}-rhythms of human brain, the problem of possible impact  of electromagnetic fields of "pearl" type onto stability of mentioned brain bio-rhythms arises.

Thus, investigation of possible direct correlation between the values of average annual frequencies  of resonant electromagnetic signals of "pearl" type appearance, which have certain impact onto brain biorhythms, and rate of average annual mortality because of diseases due to various abnormal functioning of human brain was the subject of this paper. Let us describe the basic laws of the electromagnetic fields generation and oscillations in the abovementioned resonators.

\section{Types of relaxation oscillations in Alfv\'{e}n maser}
\label{sec2}

Below we present a brief analysis of solutions for the equation describing small oscillations of the wave energy around its equilibrium state in the Alfv\'{e}n sweep maser~\citep{ref8}:
\begin{equation}
\ddot{\text{w}} + 2 v \dot{\text{w}} + \Omega_R^2 \text{w} = 0,
\label{eq1}
\end{equation}

\noindent where
\begin{align}
\text{w} &= \frac{E - E_0}{E_0}, \nonumber \\
2v &= 2 v_R \left( 1 - \frac{\gamma_0 \tau_{rec} \chi N_0}{1 + \tau_{rec}^2 \Omega_R^2} \cdot \frac{\partial}{\partial n_{is}} \ln {\left( \delta_m \gamma_0 \right)} \right)
\label{eq2}
\end{align}

\noindent and
\begin{equation}
2 v \ll \Omega_R.
\end{equation}

Here $\Omega_R$ and $2 v_R$ are characteristic frequency and oscillations decrement in Alfv\'{e}n resonator, respectively; $\text{w}$ is the oscillation of the Alfv\'{e}n wave energy with respect to its equilibrium value $E_0$. The latter is defined for the case of no changes in the Earth's ionosphere. These changes define reflection coefficients of Alfv\'{e}n waves and, consequently, their attenuation in the resonator under consideration; $\tau_{rec}$ is characteristic recombination time in ionosphere plasma; $N_0$ is total number of fast particles in Geomagnetic field line having unit cross section at the ionosphere level; $n_{is}$ is electron concentration in the ionosphere;  $\chi = ( \vec{k}\hat{,} \vec{v}_g )$, $k$ is wave vector; $v_g$ is Alfv\'{e}n waves group velocity; $\gamma_0$ corresponds to the steady state value of attenuation factor
\begin{equation}
\gamma = \left \vert \ln R(\omega) \right \vert / \tau_g,
\label{eq3}
\end{equation}

\noindent where $R(\omega)$ is a coefficient of Alfv\'{e}n wave reflection from the magnetic mirrors, that lasts over ionosphere and planet surface; $\tau_g$ is propagation time of electromagnetic signal along geomagnetic field line between magneto-conjugated regions of the ionosphere; $\delta_m = \phi(\omega_m) \delta$, $\phi(\omega_m)$ is normalized to unity Alfv\'{e}n wave amplification for one pass along the radiation belt (RB), while coefficient $\delta$ equals to:
\begin{equation}
\delta = \frac{4 \pi e^2 \beta_0}{m_e n_A \Omega_L W_0}, ~~ \beta_0 = \frac{v_0}{c}, ~~ W_0 = \frac{1}{2} m_e v_0^2, ~~ n_A = \frac{\Omega_{pL}}{\Omega_L},
\label{eq4}
\end{equation}

\noindent where $v_0$ is typical velocity of particles,  $\Omega_{pL}$ is the ion plasma frequency of background plasma in magnetosphere RB equatorial cross-section; $\Omega_L$ is gyro frequency in magnetosphere equatorial cross-section. It should be noted here, that the following approximation for coefficient $\delta_m$ is used for taking into account the effect of the generated frequency sweep (drift):
\begin{equation}
\delta_m = \delta_0 + \frac{\partial \delta_m}{\partial n_{is}} \Delta n_{is},
\label{eq5}
\end{equation}

\noindent where $\delta_0$ corresponds to equilibrium value of coefficient $\delta$, which is obtained for the case of stationary solution for differential equations which describe the dynamics of cyclotron instability in case of equilibrium electron density in the ionosphere.

It is convenient for analysis of typical forms of relaxation oscillations in Alfv\'{e}n sweep-maser to rewrite the equations (\ref{eq1}) in the following way:
\begin{equation}
\ddot{\text{w}} + \lambda \dot{\text{w}} + \Gamma \lambda \text{w} = 0,
\label{eq6}
\end{equation}

\noindent with initial conditions 
\begin{equation}
\text{w}(0) = \text{w}_0, ~~\dot{\text{w}}(0) = 0,
\end{equation}

\noindent where $\Gamma = \Omega_R^2 / 2v$, $\lambda = 2v$ and $\text{w}_0 = \left[ E(0) - E_0 \right] / E_0$ is a normalized initial energy of Alfv\'{e}n wave.

The advantages of such representation of the Alfv\'{e}n waves energy relaxation oscillations become apparent during the study of the physical reasons for the time evolution of dispersion and, consequently, the morphology of such oscillations. For instance, it is easy to show that the Polyakov-Rappoport-Trakhtengerts equation (\ref{eq1}) is equivalent to the following integro-differential equation:
\begin{equation}
\dot{\text{w}} + \left( \frac{\Omega_R^2}{2v} \right)  \cdot 2v \int \limits_0^t e^{-2v(t-t')} \text{w}(t') dt' = 0, ~~ \text{w}(0) = \text{w}_0.
\label{eq7}
\end{equation}

As follows from (\ref{eq7}), the medium memory function which characterizes its "elastic" properties, has the following form:
\begin{equation}
f(t-t') = u(t-t') \cdot 2v \cdot e^{-2v (t-t')}, ~~ \tau_\lambda = 1/2v,
\label{eq8}
\end{equation}

\noindent where $u(t-t')$ is the unit Heaviside function. Obviously, when $v \to \infty$, equation (\ref{eq8}) gains the $\delta$-asymptotycs
\begin{equation}
f(t-t') \rightarrow \delta (t-t'),
\label{eq9}
\end{equation}

\noindent while the equation (\ref{eq7}) and, consequently, equation (\ref{eq1}) as well, turns into a trivial relaxation equation of exponential type with the initial conditions:
\begin{equation}
\text{w} = \text{w}_0 e^{-\left( \Omega_R^2 / 2v \right) t} = \text{w}_0 e^{-\Gamma t},
\label{eq10}
\end{equation}

\noindent where $1 / \Gamma = 2v / \Omega_R^2$ is the time of $\text{w}$ function "viscosity" relaxation to an equilibrium value $\text{w}_0$, i.e. it is the relaxation time $\tau_M$ of the Maxwellian energy distribution $\text{w}$ to the equilibrium.

Willing to preserve the properties of "viscosity" ($\tau_M$) and "quasi-elasiticity" ($\tau_\lambda$) of the medium in equation (\ref{eq1}) let us hereinafter consider the equation (\ref{eq1}) in the form of (\ref{eq6}) with any finite $\tau_\lambda = 1 / \lambda$.

So the characteristic equation, corresponding to (\ref{eq6}) has the roots
\begin{equation}
k_{1,2} = \frac{\lambda}{2} \left( 1 \mp \sqrt{1 - 4\eta} \right), ~~ \eta = \frac{\Gamma}{\lambda} = \Gamma \tau_\lambda,
\label{eq11}
\end{equation}

\noindent which are real when $\eta < 1/4$ so that the effective time of the medium aftereffect $\tau_\lambda < \tau_M / 4$, where $\tau_M = 1 / \Gamma$ is the time of the Maxwellian energy distribution settling. Particularly, in the case of a very short medium memory ($\eta \ll 1$) we have
\begin{align}
k_1 &= \lambda \eta (1 + \eta + \dots ) \cong \Gamma (1 + \eta), \nonumber \\
k_2 &= \lambda (1 - \eta + \dots ) \cong \Gamma (1-\eta) / \eta.
\label{eq12}
\end{align}

In the case of $\eta > 1/4$, which corresponds to $\Omega_R \gg 2 v_R$, or $\tau_M < 4 \tau_\lambda$ (medium with a significant "elasticity")
\begin{equation}
k_{1,2} = \frac{1}{2} \lambda \mp i\omega, ~~ \omega = \frac{1}{2} \lambda \sqrt{4\eta - 1}.
\label{eq13}
\end{equation}

Then the general solution satisfying the initial conditions has the form
\begin{align}
&\text{w} = \text{w}_0 \frac{1}{k_1 + k_2} \left( k_2 e^{-k_1 t} - k_1 e^{-k_2 t} \right) , \nonumber \\
&\text{w}_0 = \frac{E(0) - E_0}{E_0}.
\label{eq14}
\end{align}

\begin{figure}
  \begin{center}
  \includegraphics[width=6cm]{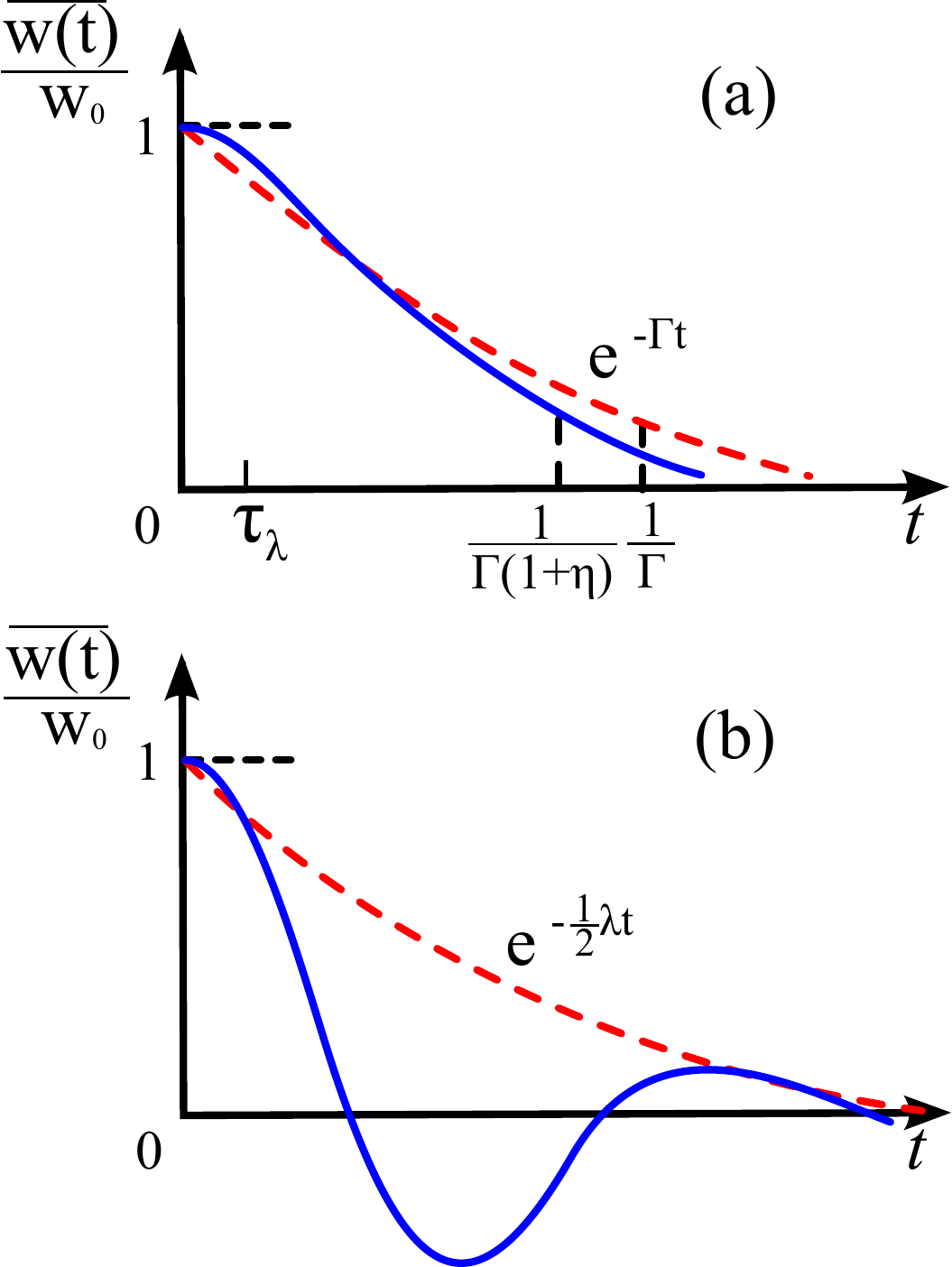}  
  \end{center}
\caption{Exponential \textbf{(a)} and oscillatory \textbf{(b)} types of relaxations of Alfv\'{e}n wave energy perturbations }
\label{fig3}
\end{figure}

In the case of $\eta < 1/4$ the solution (\ref{eq14}) takes on the following form:
\begin{equation}
\text{w} \cong \text{w}_0 \left[ (1 + \eta) e^{-\Gamma (1 + \eta) t} - \eta e^{-\lambda t} \right], ~~ \eta \ll 1,
\label{eq15}
\end{equation}

\noindent which describes the exponential relaxation which is qualitatively different from $e^{-\Gamma t}$ (a case of $\tau_\lambda = 0$) only in the range $0 < t < \tau_\lambda = 1/\lambda$ (Fig.~\ref{fig3}a). Meanwhile for the case of $\eta > 1/4$ the solution (\ref{eq14}) describes a new kind of mode

\begin{equation}
\text{w} = \text{w}_0 e^{-\lambda t /2} \frac{\sin (\omega t + \varphi)}{\sin \varphi}, ~~ \eta > \frac{1}{4}, ~~ \sin \varphi = \frac{\sqrt{4 \eta -1}}{2 \sqrt{\eta}}
\label{eq16}
\end{equation}

\noindent in a form of a attenuated periodic relaxation (Fig.~\ref{fig3}b).

Then the expression (\ref{eq16}), allowing for (\ref{eq2}) and (\ref{eq14}), may be represented in the following form
\begin{equation}
E(t) = \left[ E(0) - E_0 \right] \cdot e^{-\lambda t / 2} \frac{\sin (\omega t + \varphi )}{ \sin \varphi } + E_0, ~~ E(0) \geqslant E,
\label{eq17}
\end{equation}

\noindent where $E(t)$ is the Alfv\'{e}n waves energy.

It is known that the spectral analysis of experimental data corresponding to registration of magnetosphere radiation of "pearl" type or, in other words, geomagnetic pulsations Pc1, allows one to reveal their internal frequency structure~\citep{ref2}, and  the frequency inside of each "pearl" (separate packet of Alfv\'{e}n waves) increases from its beginning to the end~\citep{ref7}.  Quantitative estimates of "pearl" parameters following from the theory above are in a good agreement with the related experiments \citep{ref7,ref8}. Relying on that theory and experiment correspondence we will show below how the mentioned above morphological features find their explanations (within the framework of Alfv\'{e}n sweep-maser theory) on the basis of evolution of damping periodic oscillations (\ref{eq16}) or (\ref{eq17}) with taking into account dispersion relaxation of Alfv\'{e}n wave energy to its equilibrium value.

\section{Maxwell distribution and relaxation of Alfv\'{e}n wave energy dispersion toward equilibrium value}
\label{sec3}

According to Eq.~(\ref{eq14}), deviation of Alfv\'{e}n wave energy fluctuation from average value is given by
\begin{align}
&\Delta \text{w} (t) = \text{w}(t) - \overline{\text{w}(t)} = \nonumber \\
&=\int \limits_0^t \frac{1}{k_2 - k_1} \left( k_2 e^{-k_1 (t-t')} - k_1 e^{-k_2 (t-t')} \right) \xi(t') dt',
\label{eq18}
\end{align}

\noindent where, according to (\ref{eq14}) 
\begin{equation}
\overline{\text{w}(t)} = \text{w}_0 \frac{1}{k_2 - k_1} \left( k_2 e^{-k_1 t} - k_1 e^{-k_2 t} \right)
\label{eq19}
\end{equation}

\noindent and $\xi (t)$ is normalized Gaussian noise with the following moments:
\begin{equation}
\overline{\xi (t)} = 0, ~~ \overline{\xi (t) \xi (t')} = \phi (t - t') = \phi \tau \delta (t - t').
\label{eq20}
\end{equation}

Let us consider the standard deviation
\begin{align}
\overline{(\Delta \text{w} )^2} = \int \limits_0^t dt_1 \int \limits_0^t dt_2 &\prod \limits_{t=1}^2 \frac{1}{k_2 - k_1} \left( k_2 e^{-k_1 (t-t_i)} - \right. \nonumber \\
&\left. - k_1 e^{-k_2 (t-t_i)} \right) \overline{\xi (t_1) \xi (t_2)}.
\label{eq21}
\end{align}

Then, after integration (\ref{eq21}) and taking into account (\ref{eq18}) and (\ref{eq20}) we obtain the following general expression for dispersion of Alfv\'{e}n wave energy:
\begin{align}
&\overline{(\Delta \text{w})^2} = \frac{\phi \tau}{(k_2 - k_1)^2} \left[ \frac{k_2^2}{2k_1} \left( 1 - e^{-2k_1 t} \right) - \right. \nonumber \\
&\left. - \frac{2k_1 k_2}{k_1 + k_2} \left( 1 - e^{-(k_1 + k_2)t} \right) + \frac{k_1^2}{2k_2} \left( 1 - e^{-2k_2 t} \right) \right].
\label{eq22}
\end{align}

Assuming that for $t \gg \tau_M = 1/\Gamma$ the energy distribution of Alfv\'{e}n wave is relaxing to Maxwell distribution \citep{ref16,ref17}:
\begin{equation}
\left. \overline{( \Delta \text{w} )^2} \right \vert_{t \gg 1/\Gamma} = \overline{\text{w}^2} = \theta^2 \cdot \frac{\partial \overline{\text{w}}}{\partial \theta}, ~~ \theta = \frac{kT}{E_0},
\label{eq23}
\end{equation}

we have
\begin{equation}
\frac{\phi \tau}{(k_2 - k_1)^3} = \frac{\theta ^2 (\partial \overline{\text{w}}/\partial \theta)}{k_2^2 /2k_1 - 2k_1 k_2 / (k_1 + k_2) + k_1^2 / 2k_2}.
\label{eq24}
\end{equation}

\begin{figure}
\begin{center}
  \includegraphics[width=6cm]{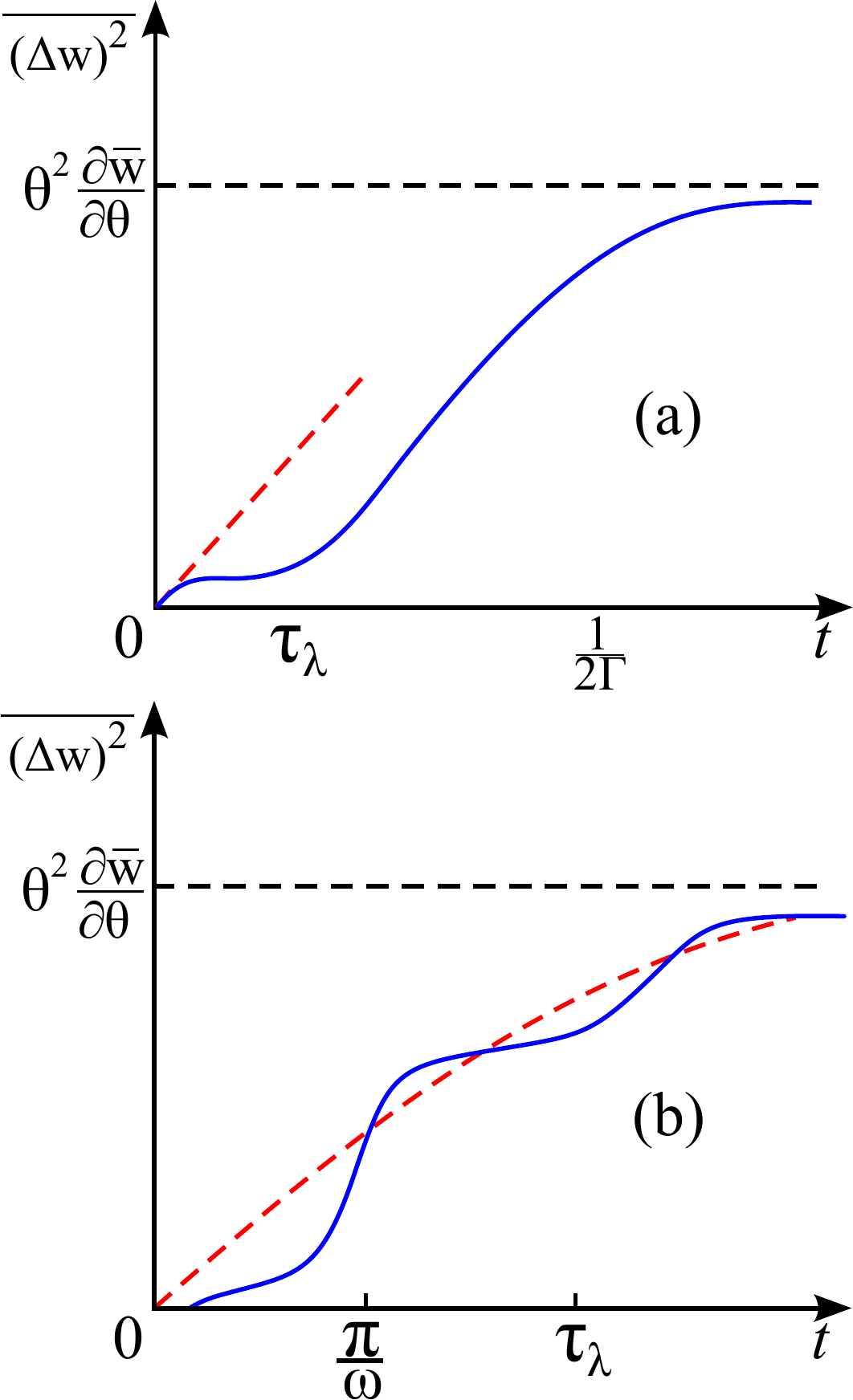}
\end{center}
\caption{Relaxation of Alfv\'{e}n wave energy dispersion $var (\text{w})$ to equilibrium value for aperiodic \textbf{(a)} and oscillatory \textbf{(b)} character of relaxation in the medium with memory.}
\label{fig4}
\end{figure}

In the case of $\eta \ll 1$ ($\tau_\lambda \ll \tau_M$) the dispersion relaxation to its equilibrium value (\ref{eq23}) is schematically shown in Fig.~\ref{fig4}a. It is characterized by three relaxation times:
\begin{equation}
\frac{1}{2k_2} = \frac{\tau_\lambda}{2} (1+\eta), ~~ \frac{1}{k_1 + k_2	} = \tau_\lambda, ~~ \frac{1}{2k_1} = \frac{1}{2\Gamma} (1-\eta).
\label{eq25}
\end{equation}

In the case of $\eta > 1/4$, when according to Eq.~(\ref{eq13}), $k_{1,2} = \lambda/2 \pm i \omega$ and the relaxation character of dispersion of Alfv\'{e}n wave energy to its equilibrium value (\ref{eq23}) becomes an oscillatory one (Fig.~\ref{fig4}b):
\begin{equation}
\overline{(\Delta \text{w} )^2} = \theta^2 \frac{\partial \overline{\text{w}}}{\partial \theta} \cdot \left[ 1 - e^{-\lambda t} \frac{1 - (1 / \sqrt{4 \eta} \cos (2 \omega t + 3 \varphi )}{1 - (1/\sqrt{4 \eta}) \cos 3 \varphi} \right],
\label{eq26}
\end{equation}

\noindent where value $\varphi$ is defined according to formula after Eq.~(\ref{eq16}) 
\begin{equation}
\varphi = \arctan \sqrt{4 \eta - 1},
\label{eq27}
\end{equation}

\noindent and the thermodynamical function $\partial \overline{\text{w}} / \partial \theta$ by definition is the thermal capacity $C_V$, which in the simplest case for plasma has the following form~\citep{ref16}:
\begin{equation}
C_V = \left( \frac{\partial \overline{\text{w}}}{\partial \theta} \right) _V = \left( C_{ideal} \right) _V + \frac{1}{2} \frac{A}{T^{3/2} V^{1/2}}, ~~ A = const,
\label{eq28}
\end{equation}

\noindent where $C_{ideal}$ is the thermal capacity of ideal gas.

Let us remind that the expression (\ref{eq26}) taking into account (\ref{eq2}) and (\ref{eq23}) can be presented in the following form:
\begin{equation}
\overline{(\Delta E )^2} = \theta^2_T \cdot \frac{\partial \overline{E}}{\partial \theta _T} \cdot \left[ 1 - e^{-\lambda t} \frac{1 - (1 / \sqrt{4 \eta} \cos (2 \omega t + 3 \varphi )}{1 - (1/\sqrt{4 \eta}) \cos 3 \varphi} \right],
\label{eq29}
\end{equation}

\noindent where $\theta_T = kT$.

\begin{figure}
\begin{center}
  \includegraphics[width=8cm]{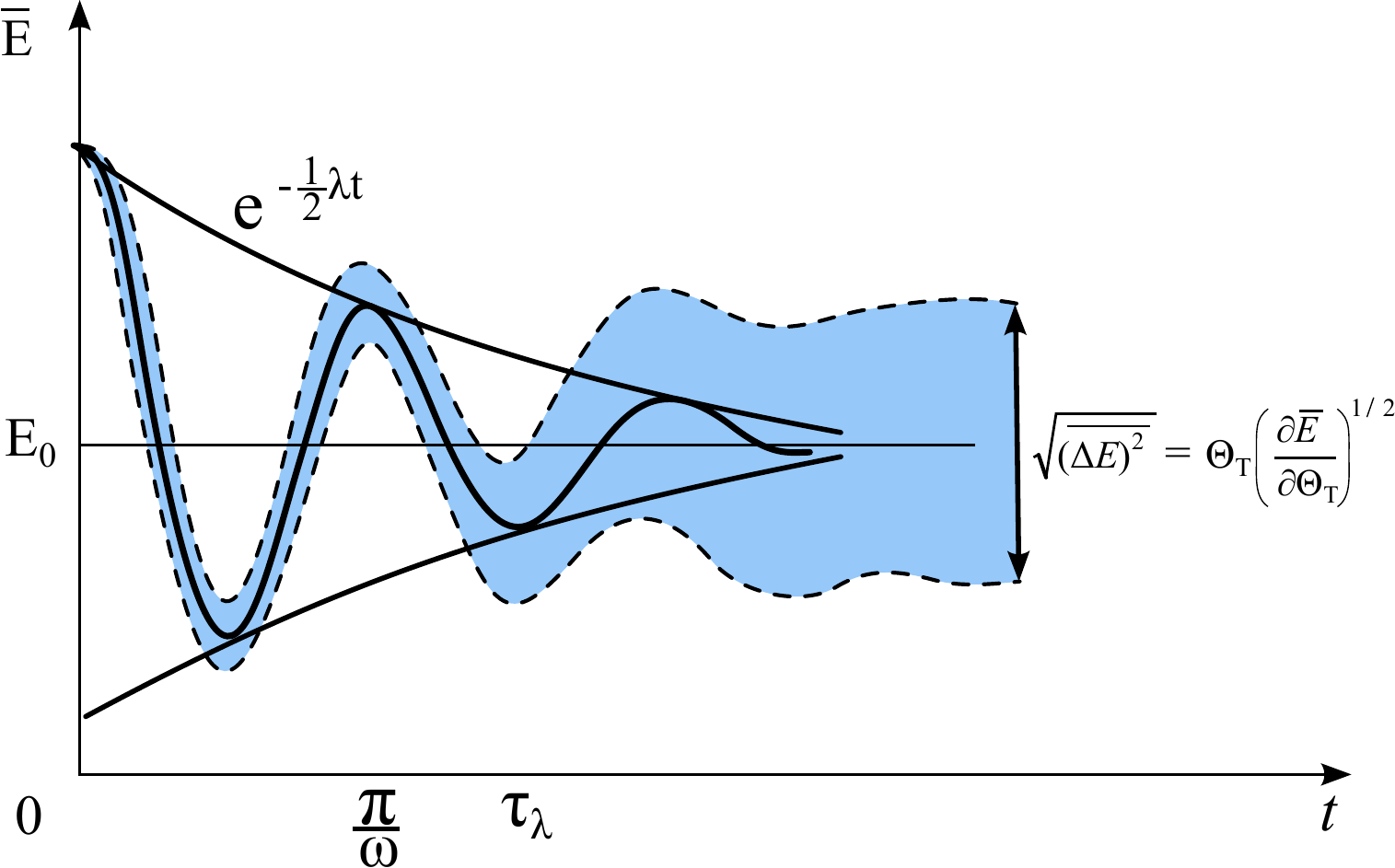}
\end{center}
\caption{Oscillatory relaxation of Alfv\'{e}n wave average energy $\bar{E}$ and its dispersion $var (E)$ in magnetoplasma medium (medium with memory).}
\label{fig5}
\end{figure}

Temporal behavior of average energy $\overline{E}$~(\ref{eq17}) and energy root-mean-square deviation $\left( \overline{\Delta E} \right)^{1/2}$~(\ref{eq29}) of relaxation oscillations of Alfv\'{e}n waves in magnetoplasma medium (medium with memory) are presented in Fig.~\ref{fig5}.

It is worth  noting here, that such approach opens up nontrivial possibility for experimental numerical estimations of some important parameters of Alfv\'{e}n sweep-maser relaxation oscillations. For instance, the limit width of distribution (\ref{eq23}) allows determining the plasma thermal capacity $C_V$ for the given temperature $\theta$. In combination with the measured frequency of oscillations $\Omega_R$ it allows successively finding the values $\lambda$(at $t = \pi / \omega$), $\eta$ (for any $t < \tau_\lambda = 1/ \lambda$) and $\tau_m = 1 / \Gamma$.

In other words, analysis of energy dispersion evolution of Alfv\'{e}n sweep-maser relaxation oscillations allows finding experimentally (see Fig.~\ref{fig5}) the values of plasma thermal capacity $C_V$, decrement $\lambda$, relaxation time ("elasticity") $\tau_\lambda$ of magnetoplasma medium and settling time ("viscosity") $\tau_M$ of Maxwellian energy distribution (see Fig.~\ref{fig3}).

And finally, folowing \citep{ref8}, let us give some quantitative estimations. Accrording to \citep{ref15}, the recurrence period of the elements in "pearls" is about 50-300~s (Fig.~\ref{fig2}), the bandwidth $\Delta f$ fits into the range 0.05-0.3~Hz, and the dynamic spectrum tilt is
\begin{equation}
\frac{df}{dt} \cong 2 \cdot 10^{-3} ~[Hz].
\label{eq30}
\end{equation}

In order to estimate the radiation parameters following from the sweep-maser theory \citep{ref8}, let us consider a magnetic flux tube on the morning side of a magnetosphere\footnote{Geomagnetic pulsations of "pearl" type are known \citep{ref15} to appear primarily on the morning side of magnetosphere at mid latitudes at magnetically calm times.} at a distance of $R \approx 3 R_{Earth}$ from the Earth center, where $R_{Earth}$ is the Earth radius. In the framework of the sweep-maser theory it has been shown that the relaxation oscillations period, which characterizes the recurrence period of the elements in pulsations of the "pearl" type is
\begin{equation}
T_R = \frac{2 \pi}{\Omega _R} = 2 \pi \sqrt{\frac{\sigma \cdot l}{W_0 \cdot \delta \cdot 2 S_0}},
\label{eq31}
\end{equation}

\noindent where $\Omega_R$ is the characteristic frequency of relaxation oscillations in Alfv\'{e}n resonator, $\sigma = B_m / B_L$ is the mirror ratio for the Earth's radiation belt, $B_m$ is the magnetic field at the ends of the magnetic trap, $B_L$ is a magnetic field in equatorial section of magnetosphere, $l$ is the effective length of a resonator, $S_0$ is the equilibrium precipitating protons flux density in the Earth's radiation belt.

According to \citep{ref8}, the amplification curve $\phi (\omega_m)$ (see.~(\ref{eq2})) reaches its maximum when
\begin{equation}
\frac{\Omega _L ^2}{\Omega _{pL} \beta_0 \omega_m} \cong 1.
\label{eq32}
\end{equation}

If we take into account the experimental values for $\Omega_L \approx 10^2 ~s^{-1}$, $n_A \approx 10$ and $\beta_0 \approx 2 \cdot 10^{-2}$ (for the particles with energy $W_0 \approx 200 ~keV$) and allow for (\ref{eq4}) and (\ref{eq32}), we find that
\begin{equation}
\omega_m \sim 5 s^{-1} \Leftrightarrow f_m \sim 1.25 ~Hz.
\label{eq33}
\end{equation}

On the other hand, from (\ref{eq4}), (\ref{eq32}) and (\ref{eq33}) it is not hard to find the value of $W_0 \cdot \delta$, which (if we assume that $n_{is L} \approx 3 \cdot 10^2 ~cm^{-3}$) would be equal
\begin{equation}
W_0 \delta \sim \frac{10 \Omega_L^2}{\omega _m n_{is L}} \approx 10^2 ~~ \left[ s^{-1} \right].
\label{eq34}
\end{equation}

\noindent where the ionospheric plasma density $n_{is L}$ is defined by the so-called plasma frequency
\begin{equation}
\Omega_{pL} = \sqrt{4 \pi e^2 n_{is L} / m_e}, \nonumber
\label{eq34a}
\end{equation}

\noindent which determines the characteristic time scale of the plasma oscillations.

Consequently, taking into account (\ref{eq31}), (\ref{eq34}) and the experimental data for $\sigma = 27$ and $l \approx R$ we derive
\begin{equation}
T_R = \frac{2 \pi}{\Omega _R} = 2 \pi \frac{10^4}{S_0 ^{1/2}}.
\label{eq35}
\end{equation}

The period $T_R$ obviously hits the experimentally observed range $50 \div 300 ~s$ with the experimentally justified value  $S_0 \sim 10^5 ~cm^{-2}s^{-1}$.

Let us pass on to the dynamic spectrum tilt estimation:
\begin{equation}
\frac{df}{dt} \approx \frac{df}{dn_{is}} \cdot \frac{dn_{is}}{dt} \cong \chi \cdot S_0 \cdot \frac{df}{dn_{is}},
\label{eq36}
\end{equation}

\noindent where $n_{is}$ is the ionospheric plasma density, $cm^{-3}$.

According to \citep{ref8}, the numerical calculation of $df / d n_{is}$ for the calm morning gives the value $\sim 3 \cdot 10^{-6} ~cm^3 \cdot s^{-1}$. On the other hand there are reasons to believe that under weak magnetic storminess the protons with energy about 200~keV (and $\chi \approx 10^{-2} ~cm^{-1}$) flux density is $S_0 \sim 10^5 ~cm^{-2} s^{-1}$. From this it follows that the dynamic spectrum tilt is
\begin{equation}
\frac{df}{dt} \approx \chi \cdot S_0 \cdot \frac{df}{dn_{is}} \sim 3 \cdot 10^{-3} ~~\left[ Hz \cdot s^{-1} \right],
\label{eq37}
\end{equation}

\noindent which corresponds to the experimental data \citep{ref15} with a satisfiable accuracy.

Therefore we may conclude that the Polyakov-Rappoport-Trakhtengerts sweep-maser theory \citep{ref6,ref7,ref8,ref9,ref10,ref11,ref12} makes it possible to build a closed theory of generation of a wide range of geomagnetic pulsations of the "pearl" type, which reside in the 0.1$\div$10~Hz band and are observed primarily at the mid latitudes under the conditions of a weak magnetic storminess on the morning side of magnetosphere. In other words, it is shown that all the morphological features of geomagnetic pulsations of the "pearl" type mentioned above find their adequate explanation in the framework of the Alfv\'{e}n sweep-maser theory.

\section{On connections between variations of geomagnetic pulsations, solar cycles and brain diseases mortality rate}
\label{sec4}

\begin{figure}
\begin{center}
  \includegraphics[width=6cm]{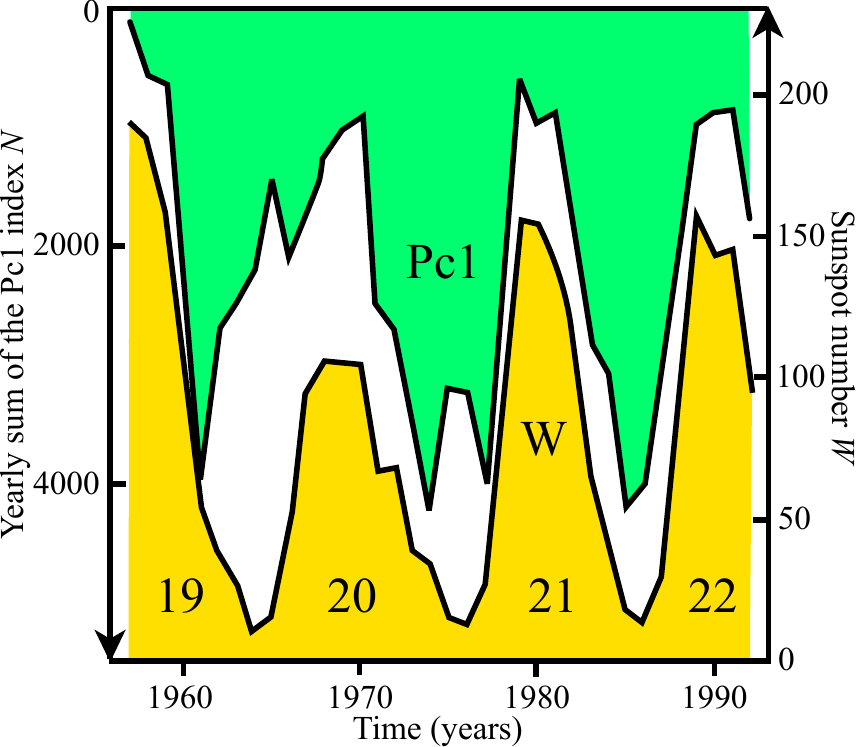}
\end{center}
\caption{Solar cycle (yellow) variations of geomagnetic pulsations Pc1 (green) activity for about four (from 19$^{th}$ through 21$^{st}$) cycles \citep{ref15}.}
\label{fig6}
\end{figure}

As mentioned above, the theory of formation of the "pearl" wave packets (geomagnetic pulsations Pc1) in Alfv\'{e}n maser may help in solving another  problem. This problem lies in the fact of strong \textit{inverse} correlation between the appearance frequency of geomagnetic pulsations Pc1 and 11-year solar cycle. It has been found by means of long-term observations that geomagnetic pulsations Pc1 activity is more intense (by factor of 10) during the periods of solar minima rather than in its maxima (Fig.\ref{fig6}). Below we will try to clarify this dependency within the frame of Alfv\'{e}n maser theory.

\begin{figure}
\begin{center}
  \includegraphics[width=6cm]{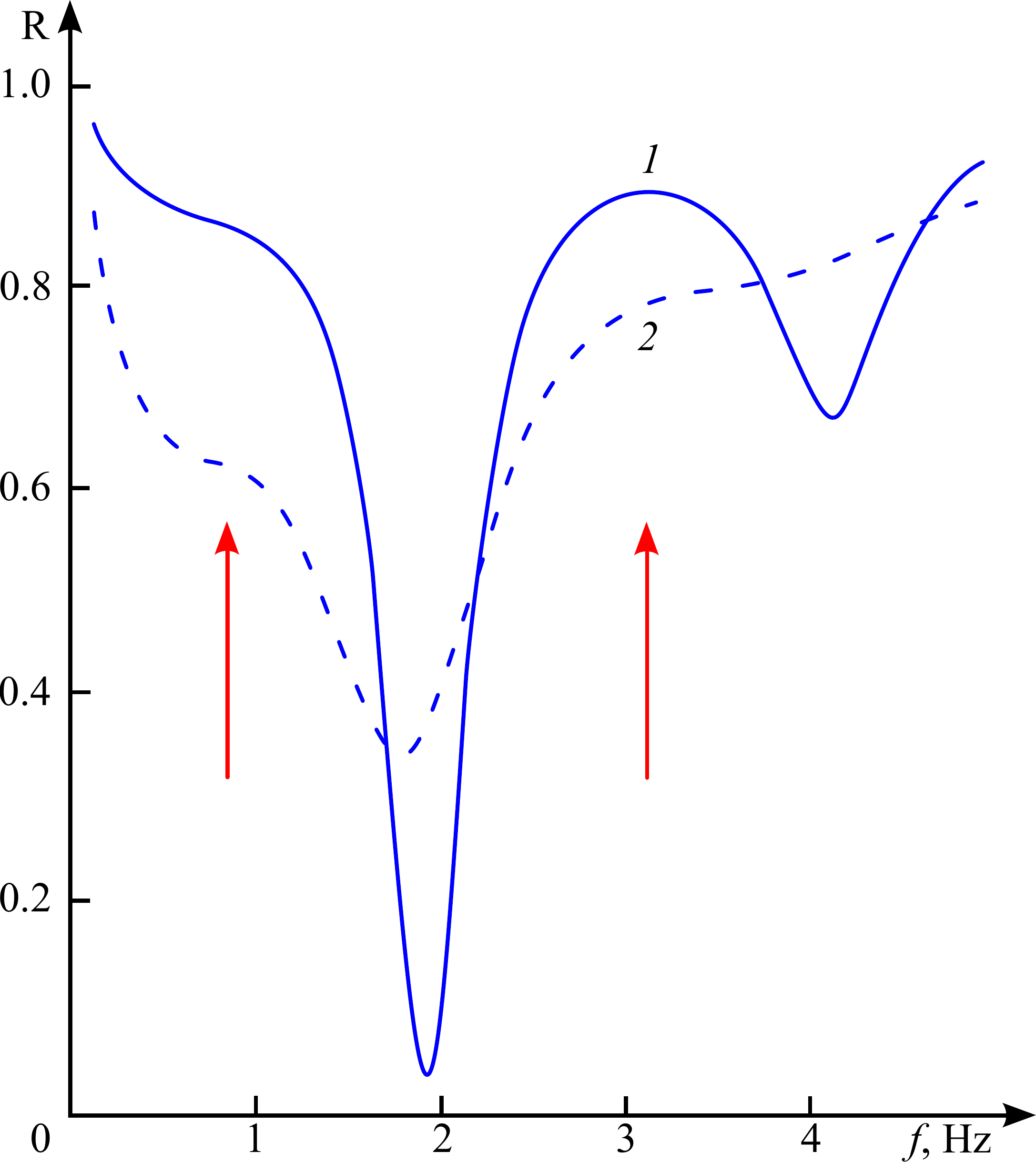}
\end{center}
\caption{Frequency dependence of Alfv\'{e}n wave reflection coefficient R from ionosphere containing IAR \citep{ref7}: 1 - for solar activity minimum, 2 - for solar activity maxim. Drop of plasma concentration at altitudes of 250-1000~km is much more pronounced in the solar activity minimum. It leads to the greater value of reflection coefficient $R$ (regions of $R$ maximal values are indicated with arrows), where high rate generation of "pearls" takes place.}
\label{fig7}
\end{figure}

Experimental observations of ionosphere have shown that steepness of electron concentration profile at the attitudes $\sim$1000~km (see Fig.~\ref{fig1}b) is considerably decreasing in the years of solar activity maxima \citep{ref18}. This factor leads to decrease in the Alfv\'{e}n waves reflection coefficient from the upper layer of the IAR and, hence, to decrease in Q-factor of AR. Fig.~\ref{fig7} shows experimental dependence of the reflection coefficient $R$ of Alfv\'{e}n waves from IAR. It was obtained using ionosphere data for the minimum and maximum of solar activity \citep{ref18}. One can see that appreciable decrease in the reflection coefficient R (and, therefore, worsening of the conditions for "pearl" generation in magnetospheric Alfv\'{e}n resonator (AR)) is observed for maximum of solar activity compared to its minimum. The explanation of such behavior of the reflection coefficient $R$ and, consequently, behavior of the "pearl" generation rate is rather straightforward and is given below.

The criterion for the wave generation in Alfv\'{e}n maser according to (\ref{eq3}) has the form \citep{ref6,ref7}
\begin{equation}
R(\omega) \cdot \exp \Gamma_0 > 1,
\label{eq38}
\end{equation}

\noindent where $\Gamma_0 = \gamma \tau_g$ is the logarithmic wave amplification at a single passing of AR (Fig.~\ref{fig1}a), $R(\omega)$ is a coefficient of Alfv\'{e}n wave reflection from the magnetic mirrors, that lasts over ionosphere and planet surface. The "pearl" amplification changes little during the solar activity cycle and has its value considerably less than unity. Whereas the maximal value of reflection coefficient $R$ in the typical for the "pearls" frequency band of 0.2-5~Hz changes considerably according to Fig.~\ref{fig7} and decreases in the years of solar activity maxima. So, it is follows from the (\ref{eq38}) that the temporal variations of IAR Q-factor affect the appearance rate of "pearls" generation. In other words, reflection coefficient R behavior clearly explains (via the criterion of wave generation in Alfv\'{e}n maser (\ref{eq38})) the dynamics of "pearls" appearance rate, which, in turn, explains the reason for strong anticorrelation between "pearls" appearance and solar activity (Fig.~\ref{fig6}).

\begin{figure*}
\begin{center}
  \includegraphics[width=16cm]{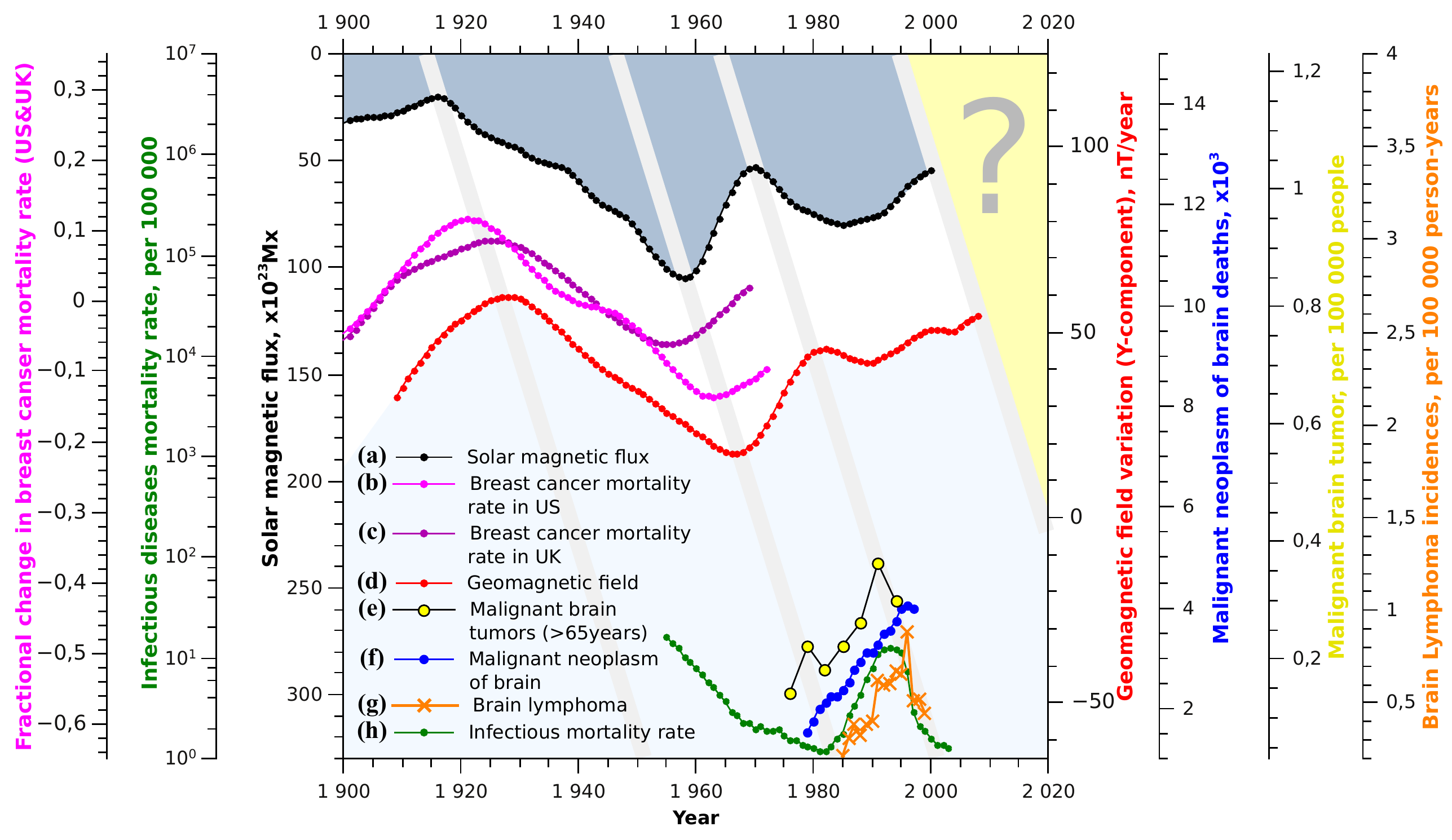}
\end{center}
\caption{Time evolution \textbf{(a)} the variations of magnetic flux at the bottom (tachocline zone) of the Sun convective zone (see Fig.~7f in~\citep{ref24}), \textbf{(b)} fractional change in female breast cancer mortality for birth cohort in US (see Fig.~3b in~\citep{Juckett2009}), \textbf{(c)} fractional change in female breast cancer mortality for birth cohort in UK (see Fig.~2b in~\citep{Juckett2009}), \textbf{(d)} geomagnetic field secular variations (Y-component, nT/year) as observed at the Eskdalemuir observatory (England) \citep{ref25}, where the variations ($\partial Y / \partial t$) are directly proportional to the westward drift of magnetic features, \textbf{(e)} Malignant brain tumor (brain stem) \citep{Legler1999}, \textbf{(f)} the number of deaths from ICD9 item n$^{\circ}$191  Malignant neoplasm of brain \citep{ref21}, \textbf{(g)} Brain lymphoma incidences in US \citep{Hoffman2006} and \textbf{(h)} the mortality rates from infectious diseases (incl. Tuberculosis) at ages 15-34 in France \citep{ref23}. The curves \textbf{(a}) and \textbf{(d)} are smoothed by the sliding intervals of 5 and 11 years.}
\label{fig8}
\end{figure*}

Using this dependence and having temporal evolution of solar activity or, that the same, the temporal evolution of Sun's magnetic field we may conclude about our principal knowledge of temporal evolution of "pearls" appearance over the past 100 years, at the least. Temporal dynamics of Sun's magnetic field is depicted in Fig.~\ref{fig8}. Existence of strict anticorrelation between Sun's magnetic field and terrestrial magnetic field\footnote{Note that the strong (inverse) correlation between the temporal variations of magnetic flux in the tachocline zone and the Earth magnetic field (Y-component) are observed only for experimental data obtained at that observatories where the temporal variations of declination ($\partial D / \partial t$) or the closely associated east component  ($\partial Y / \partial t$) are directly proportional to the westward drift of magnetic features \citep{ref19}. This condition is very important for understanding of physical nature of indicated above correlation, so far as it is known that just motions of the top layers of the Earth's core are responsible for most magnetic variations and, in particular, for the westward drift of magnetic features seen on the Earth's surface on the decade time scale. Europe and Australia are geographical places, where this condition is fulfilled (see Fig.~2 in \citep{ref19}).} (Y-component)\citep{ref20} is seen from this figure, as well.

Due to the fact that the frequency band of Alfv\'{e}n maser resonant structures practically coincides with the frequency band of \textit{delta}-rhythms and, partially, \textit{theta}-rhythms of human brain (see Fig.~\ref{fig7}), the question naturally arises about the rate of possible influence of global geomagnetic pulsations of "pearl" type onto stability of the above brain biorhythms. If such effect really exists, then one would expect positive correlation between variation of geomagnetic pulsations of "pearl" type Pc1 in the frequency band of 0.1-5~Hz and the death rate from disruption of brain diseases. Obviously, the choice in this case should concern only the currently incurable brain diseases so that their statistics were close to the real one and not being masked by intensive treatment. This fully applies to such diseases as malignant neoplasm of brain \citep{ref21}. Their temporal dynamics in West Germany \citep{ref21} is shown in Fig.~\ref{fig8}. It is of interest that for our purposes the male statistics of infectious diseases (incl. Tuberculosis) in France \citep{ref23} also applicable, that reflects, apparently, features of local spatial-temporal dynamics of magnetic field in Europe.

It is easy to show, that degree of anti-correlation between temporal variations of Sun's magnetic field or, that the same, degree of direct correlation between frequency of "pearls" appearance and the number of deaths from considered diseases is high enough. This result is based on the experimental data on malignant neoplasm of brain \citep{ref21}, malignant brain tumor \citep{Legler1999}, brain lymphomas \citep{Hoffman2006} and infectious diseases (incl. Tuberculosis) \citep{ref23}. In this way, according to Fig.~\ref{fig8}, time arranged statistics of these diseases are lagging behind the variations of the Solar and Earth magnetic fields 22-27 years and 10-15 years respectively. This delay effect on the one hand can be a consequence of the long-time hidden disease incubation period, but on the other hand opens up possibilities for prediction (at the time lag length) of behavior variations of indicated diseases by means of experimental observation of the geomagnetic field temporal variations.

It is interesting to note here that a strong correlation between the galactic cosmic ray variations and cancer mortality birth cohorts has been discovered recently \citep{Juckett2007,ref27}. It was observed for population cohorts in five countries on the three continents. Previous evidence  \citep{ref27}  has implicated a role for cosmic rays in US female cancer, involving a possible cross-generational foetal effect (grandmother effects). According to the assumption of the authors \citep{Juckett2007,ref27}, it may provide in-sight into the exploration of the role of germ cells as a possible target of this radiation and genetic or epigenetic sources of cancer predisposition that could be used to identify individuals carrying the radiation damage. And the conclusion about the galactic cosmic rays as a direct physical cause of the cancer mortality birth cohorts is based on a similar dependence of the total cancer age-standardized incidence rates and cosmic ray rigidity from geomagnetic latitude (see Fig.~8 in~\citep{Juckett2007}).

Not reducing the role of the physical mechanisms of radiation-induced effects formation and non-linear cell response in low doses of ionizing radiation (e.g. \citep{ref28}),the study of which is a fundamental basis for the contemporary microdosimetry \citep{ref29}, let us consider the possibility of indirect impact of electromagnetic resonance structures (in 0.1-5 Hz band) of ionospheric Alfv\'{e}n maser on the cells of the birth cohorts through their direct impact on the germ cells of their parents. Fig.~\ref{fig8}b,c shows a high level of inverse correlation between the temporal variations of the solar magnetic field (or direct correlation between the "pearls" appearance frequency) and cancer mortality rate of birth cohorts.

Time lag between the inverse solar magnetic field and cancer mortality birth cohorts is $\sim$6 years for UK-data and $\sim$10 years for USA-data, as follows from Fig.~\ref{fig8}b,c. At first sight it may seem to contradict the 28-year lag between the galactic cosmic rays variations and cancer mortality birth cohorts, established in the paper by \citep{Juckett2007} basing on the data \citep{ref30,ref31,ref32}. However, it may be explained by the known and hard-to-remove effect of the time shift in ice core data accompanying any ${}^{10}$Be measurements (proxy of galactic cosmic rays) in ice cores of Greenland and Antarctica.On the other hand, theoretical verification of the actual ${}^{10}$Be-data \citep{ref33} and their comparison with the analogous data obtained from \citep{ref30,ref31,ref32} indicates that the time lag between the galactic cosmic rays variations and cancer mortality birth cohorts is $\sim$6-10~years.

\begin{figure}
\begin{center}
  \includegraphics[width=8cm]{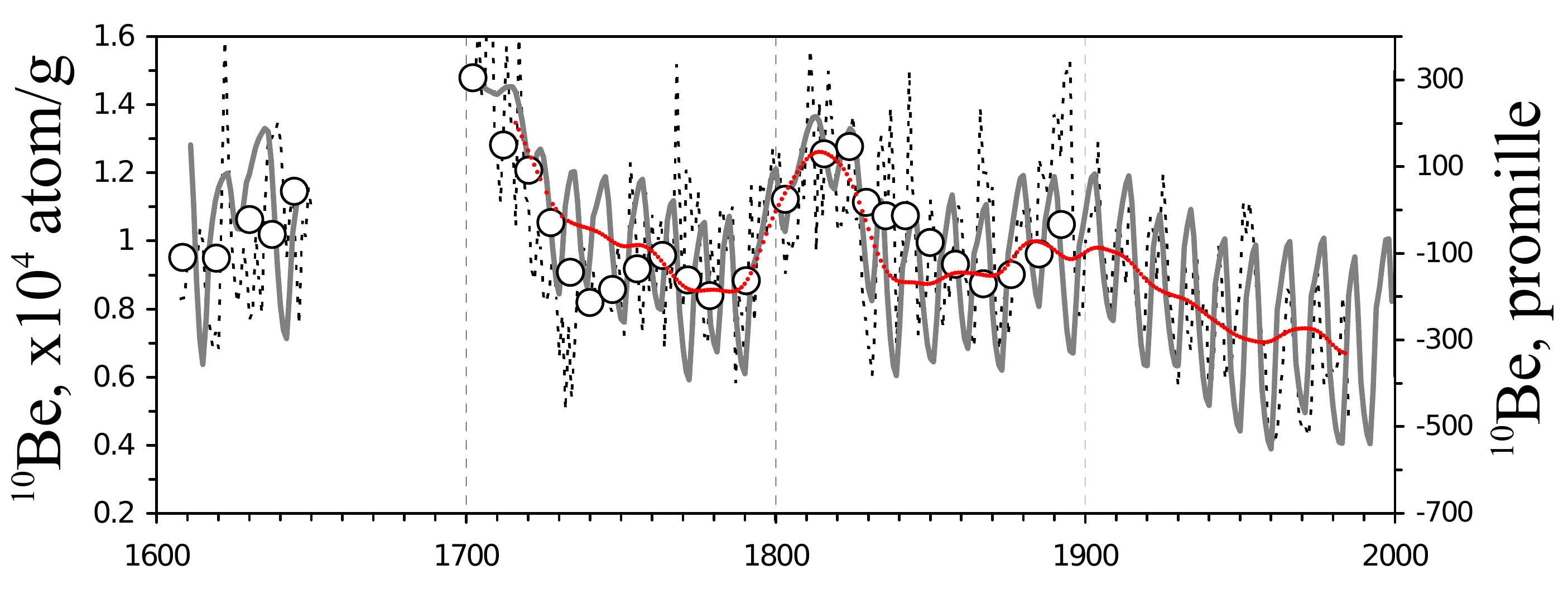}
\end{center}
\caption{Long-term cosmic rays reconstruction (from  \citep{ref33}). Calculated (grey curve) and actual annual ${}^{10}$Be content in Greenland ice (dotted curve). Open circles represent the 8-year data from Antarctica \citep{ref32}. Red line represents the 33-year moving average of the grey curve \citep{ref33}.}
\label{fig9}
\end{figure}

There is also another more trivial justification of the 10-year time lag on the Fig.~\ref{fig8}c. The variations of galactic cosmic rays are obviously a consequence of their modulation by the solar magnetic field. It means that magnetic fields of the solar wind deflect the primary flux of charged cosmic particles, which leads to a reduction of cosmogenic nuclide (e.g. ${}^{10}$Be and ${}^{14}$C) production in the Earth's atmosphere. In other words, cosmogenic nuclides (e.g. ${}^{10}$Be and ${}^{14}$C) are a kind of a "shadow" of galactic cosmic rays on the Earth playing the role of a proxy for the solar magnetic variability. Therefore the variations of the solar magnetic field and  galactic cosmic rays (or ${}^{10}$Be-proxy) must inversely coincide which visually demonstrates the experimentally justified result of the 1-year lag between ${}^{10}$Be and sunspot originally detected by Beer et al.  \citep{ref30}. After all, one could not expect anything else because the galactic cosmic rays variations are caused by the solar magnetic field variations and not vice versa.

Turning back to a direct physical cause of the cancer mortality birth cohorts, it should be noted that it is practically impossible to separate the possible radiative effect on germ cells (according to \citep{Juckett2009}) from the magnetobiological effect induced by such electromagnetic radiation as "pearls", since:
\begin{enumerate}
\item[a)] electromagnetic radiation of the "pearls" type is generated as a result of the protons (a dominant component of the cosmic rays) passage through the Alfv\'{e}n maser resonator, which is a magnetospheric magnetic flux tube resting upon the parts of ionosphere in the conjugate hemispheres of the Earth;

\item[b)] the intensity variations of electromagnetic radiation of the "pearls" type -- because of their origin -- not only is correlated with the galactic cosmic rays variations, but also display a similar latitude dependence;

\item[c)] magnetic field of the "pearls" can freely penetrate the human body just like any other magnetic field, because the human body tissues almost do not decrease their intensity; indeed, the harmonic amplitude of the field with frequency $\omega$ in a oscillatory circuit on the depth $h$ inside the body is decreased $f_s$ times
\end{enumerate}
\begin{equation}
f_s (\omega, h, \sigma, \mu ) = \exp (-h/\delta),
\label{eq39}
\end{equation}

\noindent where the path till absorption $\delta$ depends, according to \citep{Jackson1975}, on the permeability $\mu ~(\sim 1)$ and conductivity $\sigma$, and is defined as follows: $\delta = c (2 \pi \mu \omega \sigma)^{-1/2}$. Since for $\omega < 10^6 ~s^{-1}$ we have $\delta > 10^3 ~cm$, for $h \leqslant 10 ~cm$ from (\ref{eq39}) we obtain the value $f_s \sim 1$.

Taking into account the stated properties and the known fact (e.g. \citep{Nakagawa1997,Simon1992}) that the magnetic field may be a kind of an agent that amplifies the original cause (chemical impact or exposure to ionizing radiation) of the carcinogenesis, we may assume that the magnetobiological effect induced by the electromagnetic radiation of the "pearls" type amplifies the cosmic rays radiative effect in the germ cells of the parents \citep{Juckett2009}.

As one can easily see, the level of inverse correlation between the solar magnetic field variations (or direct correlation between the frequency of the "pearls" appearance) and cancer mortality rate of birth cohorts is high enough. Also, according to Fig.~\ref{fig8}c, the time series of this effect lag approximately 10 years behind the temporal variations of the Earth magnetic field. On the one hand, such delay effect may be a consequence of the cross-generational foetal effect \citep{Juckett2009}, and on the other hand, it makes it possible to predict the discussed variations by experimental observation of the geomagnetic field variations.

\section{Conclusions}
\label{sec5}

In the frames of Alfv\'{e}n maser theory the description of morphological features of relaxation oscillations in the mode of geomagnetic pulsations of "pearl" type (Pc1) in the ionosphere is obtained. These features are determined by "viscosity" and "elasticity" of magnetoplasma medium that control the nonlinear relaxation of energy and dispersion of Alfv\'{e}n wave energy to the equilibrium values.

On the basis of analysis of the "pearls" generation criterion in Alfv\'{e}n maser (\ref{eq38}) the physical reasons for strong anticorrelation of the appearance of "pearls" relatively to solar activity are discussed. Obviously, that a priori knowledge of the temporal evolution of solar activity or, that the same, the temporal evolution of Sun's magnetic field gives possibility to build the temporal evolution of "pearls" appearance over the past 100 years, at least. The latter opens up a possibility for studying of positive correlation between "pearls" appearance rate and temporal variations of death rate from various disruptions of brain diseases. The anti-correlation rate between temporal variations of Sun's magnetic field or, that the same, direct correlation rate between of "pearls" appearance rate and the number of deaths from considered diseases was demonstrated to have the high enough value. This result is supported by the experimental data on malignant neoplasm of brain \citep{ref21}, malignant brain tumor \citep{Legler1999}, brain lymphomas \citep{Hoffman2006} and infectious diseases (incl. Tuberculosis) \citep{ref23}.

The analysis of the known correlation between the galactic cosmic rays variations and  cancer mortality birth cohorts observed for population cohorts in five countries on the three continents \citep{Juckett2009} let us suggest a hypothesis of a cooperative action of the cosmic rays and electromagnetic radiation of the "pearls" type on the germ cells of the parents which is responsible for the so-called cross-generational foetal effect \citep{Juckett2009} with a lag of $\sim$6-10~years.

In conclusion, we have obtained results clearly showing the possible impact of electromagnetic resonant radiations generated in ionospheric Alfv\'{e}n maser onto stability of human brain biorhythms, such as \textit{delta}-rhythms and, partially, \textit{theta}-rhythms.


\bibliographystyle{unsrtnat}
\bibliography{PearlsBiorhythm}   

\end{document}